\documentclass[aps,prl,twocolumn,nofootinbib,floatfix,superscriptaddress ]{revtex4-2}

\usepackage{graphicx}% Include figure files
\usepackage{dcolumn}% Align table columns on decimal point
\usepackage{bm}% bold math
\usepackage[english]{babel}

\usepackage{multirow}
\usepackage{amssymb}
\usepackage{amsmath,bbm}
\usepackage[figuresright]{rotating}
\usepackage{capt-of}
\usepackage{xcolor}
\usepackage{booktabs}
\usepackage{siunitx}
\usepackage{placeins}
\usepackage{physics}
\usepackage[hidelinks]{hyperref}

\usepackage[english]{babel}
\usepackage[autolanguage]{numprint}

\newcommand{\nuc}[2]{\ensuremath{{}^{#1}\mathrm{#2}}}
\begin{document}
\title{Radii of Sn Isotopes near the Proton Dripline\\[0.2em]from Nuclear Lattice Effective Field Theory}

\author{Fabian Hildenbrand}
\email{f.hildenbrand@fz-juelich.de}
\affiliation{Institute for Advanced Simulation (IAS-4), Forschungszentrum J\"{u}lich, D-52425 J\"{u}lich, Germany}
\affiliation{Helmholtz-Institut f\"{u}r Strahlen- und Kernphysik, Bethe Center for Theoretical Physics\\
and Cluster of Excellence ``Color meets Flavor'', Universit\"{a}t Bonn, D-53115 Bonn, Germany}
\author{Serdar Elhatisari}
\email{selhatisari@gmail.com}
\affiliation{Faculty of Natural Sciences and Engineering, Gaziantep Islam Science and Technology University, Gaziantep 27010, Turkey}

\author{Ulf-G.~Mei{\ss}ner}
\email{meissner@hiskp.uni-bonn.de}
\affiliation{Helmholtz-Institut f\"{u}r Strahlen- und Kernphysik, Bethe Center for Theoretical Physics\\
and Cluster of Excellence ``Color meets Flavor'', Universit\"{a}t Bonn, D-53115 Bonn, Germany}
\affiliation{Institute for Advanced Simulation (IAS-4), Forschungszentrum J\"{u}lich, D-52425 J\"{u}lich, Germany}
%\affiliation{Peng Huanwu Collaborative Center for Research and Education, International Institute for Interdisciplinary and Frontiers, Beihang %University, Beijing 100191, China}

\date{\today}

\begin{abstract}
We present the first {\em ab initio} lattice calculations of  charge and matter radii of tin isotopes $^{96}$Sn, $^{98}$Sn, $^{100}$Sn, $^{102}$Sn and $^{104}$Sn near the proton dripline using nuclear lattice effective field theory with high-fidelity two- and three-nucleon forces combined with the partial-pinhole algorithm. The resulting radii are larger then predicted from other ab-initio methods. We find that variations of the three-nucleon interaction reveal substantial radius differences even among interactions with similar binding energies. We intensively discuss of the role of different choices of the three-body forces.
\end{abstract}

%%%%%%%%%%%%%%%%%%%%%%%%%%%%%%%%%%%%

\maketitle
\section{Introduction}

The tin isotopic chain is one of the most fascinating regions of the nuclear chart and provides a particularly demanding test of nuclear interactions. It features an abundance of stable nuclei, and its closed proton shell connects a long sequence of neutron-rich nuclei to the
proton-rich region around doubly magic ${}^{100}$Sn~\cite{Faestermann:2013ng}. In contrast, the neutron-deficient end of the chain terminates rather abruptly. Here, binding energies test the balance of attraction and repulsion, whereas radii probe the spatial distribution generated by that balance. Describing both observables is therefore a more restrictive test than reproducing nuclear masses alone~\cite{Lapoux:2016exf}.

On the experimental side, binding energies are known for many tin isotopes.
Direct determinations of binding energies, however,  end at $^{100}$Sn. 
For a general overview of experimental and theoretical efforts in this region of the nuclear chart, see Ref.~\cite{Faestermann:2013ng}.
In recent years, experiments at the ISOLTRAP (CERN) and LEBIT (FRIB) facilities have provided increasingly precise binding energies and constraints on the proton drip line in this region~\cite{Nies:2024gim,Ireland:2024azu,Ireland:2025tui,Lange:2026nrq}.  Charge-radius measurements reach $^{104}$Sn~\cite{Gustafsson:2025xpd}, while the four additional isotopes studied here, $^{96,98,100,102}$Sn, remain targets for radius measurements. On the theory side, coupled-cluster (CC) and valence-space in-medium similarity renormalization group (IMSRG) calculations established access to the lightest tin isotopes~\cite{Morris:2017vxi}. Self-consistent Green's function calculations addressed their charge distributions~\cite{Arthuis:2020toz}, while Bogoliubov coupled-cluster calculations (BCC) extended the description of open-shell tin nuclei~\cite{Tichai:2023epe,Vernik:2026tvi,Demol:2026glc}. Recently, charge-radius calculations using interactions of the EM(1.8/2.0) type became available~\cite{Demol:2026glc}. Furthermore, single-reference IMSRG results are available for selected tin isotopes~\cite{Hu:2025cjl}. These developments make the region a testing ground for the interaction dependence of nuclear size.

Nuclear lattice effective field theory (NLEFT) solves the many-body problem by Euclidean-time projection on a four-dimensional lattice with $L$ sites in each spatial direction and $L_t$ temporal sites~\cite{Lahde:2019npb,Lee:2025req}. Wavefunction matching (WFM) brings a realistic interaction closer to a Hamiltonian that can be simulated with reduced sign oscillations~\cite{Elhatisari:2022zrb}. With the advent of exascale computing we were able to apply this framework to tin binding energies at $A\sim100$~\cite{Hildenbrand:2025voq}. Here we extend it to charge and matter radii using the partial-pinhole method~\cite{Ren:2025vpe} and examine the dependence on recently constructed three-nucleon interactions~\cite{Elhatisari:2026dzq}.

Radius calculations present several additional challenges. First, they require intrinsic observables free of center-of-mass motion. This is achieved using the pinhole algorithm, which samples particle positions through density insertions~\cite{Elhatisari:2017eno}. The procedure introduces an additional phase factor that reduces the signal-to-noise ratio. Second, radii are sensitive to long-distance components of the wave function whose convergence need not track that of the energy. Another challenge is the high computational cost. Radius measurements require evaluating correlations between the density operators and interaction insertions throughout the Euclidean-time evolution. Moreover, for heavy nuclei, particularly Sn, contributions from three-nucleon forces and higher-order operators converge only slowly in Euclidean time $\tau=a_tL_t$. Our aim is to determine how strongly radius observables constrain the interaction beyond binding energies. We consider separate results for individual interactions, energy-conditioned radius estimates, and an ensemble analysis, because these procedures answer different questions.

In this paper we focus on even-even isotopes $^{96}$Sn, $^{98}$Sn, $^{100}$Sn, $^{102}$Sn and $^{104}$Sn around  the doubly magic $^{100}$Sn. We calculate charge and matter radii for these isotopes and revisit the separation energies already considered in~\cite{Hildenbrand:2025voq}.
A major development in this work is the systematic investigation of recently constructed three-nucleon forces on the lattice~\cite{Elhatisari:2026dzq}. This, on the one side, leads to a linear correlation between the binding energies and radii of the considered Sn isotopes, and on the  other side allows for an improved determination of the theoretical uncertainties. The present calculations extend the WFM radius studies of Ref.~\cite{Elhatisari:2022zrb}, which reached $^{58}$Ni, to the region of $A \sim 100$.
%These calculations are enabled by the available exascale computing capabilities of JUPITER.

\section{Methodology}

We use a spatial lattice spacing $a=1.32$~fm and a temporal lattice spacing $a_t=0.20$~fm. Since radii are more sensitive to the tail of the wave function than energies, the main calculations employ $L=14$ sites per direction, corresponding to a box length $La=18.48$~fm. The box half-length is 9.24~fm, roughly twice the expected radius. The two-nucleon interaction is the WFM interaction constructed at N$^3$LO~\cite{Elhatisari:2022zrb}, while its three-nucleon sector is varied as described below.

The partial-pinhole algorithm samples a subset of particles to evaluate density correlations~\cite{Ren:2025vpe}. Compared with standard pinhole calculations~\cite{Elhatisari:2017eno}, this method improves the calculation of structural properties, such as charge and matter radii, through two-body correlation functions~\cite{Ren:2025vpe}, called
$G_{pp}$, $G_{np}$ and $G_{nn}$. 
The charge ($r_{\rm ch}$) and matter ($r_m$) radii are then obtained from~\cite{Ren:2025vpe},
\begin{subequations}
\begin{align}
  r_{\rm ch}^2 =& \langle r_p^2 \rangle + R_p^2 + \frac{N}{Z} R_n^2 + \frac{3}{4m_p^2}\,,\\
\langle r_m^2\rangle =& \frac{1}{2A^2} G_{pp} + \frac{1}{A^2} G_{pn} + \frac{1}{2A^2} G_{nn}\,,\\
\langle r_p^2\rangle =& \frac{N}{ZA^2} G_{pp} + \frac{N-Z}{ZA^2} G_{pn} - \frac{1}{A^2}G_{nn}\,.
\end{align}
\end{subequations}
The charge radius is obtained by including finite-nucleon-size corrections and the Darwin--Foldy correction, with $R_p^2=0.7056$~fm$^2$~\cite{Antognini:2013txn,Lin:2021xrc}
and $R_n^2=-0.105$~fm$^2$~\cite{Filin:2020tcs}. The proton mass is set to $m_p=938.27$~MeV. $A, Z, N$ denote the total particle, proton and neutron number, respectively. We do not include spin-orbit or additional relativistic corrections.
For a full derivation and further details of the partial-pinhole calculation, see Ref.~\cite{Ren:2025vpe}.
We did not observe any residual dependence on the number of pinned particles in the tests performed and therefore chose 12 particles, which provides a favorable balance between sign oscillations and the information gained per sample

Since our previous proof-of-principle study showed that modifications to three-body forces fitted to light- and medium-mass nuclei are needed~\cite{Hildenbrand:2025voq}, we perform a detailed analysis of the best fits of the recently developed partial-wave-resolved (PWD) three-body forces~\cite{Elhatisari:2026dzq}.
These forces are divided into two categories: one includes additional cluster-type three-body forces $(T^p_1)$, while the other does not $(T^p_2)$. Here, $p$ is the number of fitted short-range three-nucleon terms.
Both categories describe nuclear binding in the light- and medium-mass region, together with low-energy neutron-alpha scattering~\cite{Elhatisari:2025fyu}.
The additional operators in the $T^p_1$ category are motivated by cluster EFT~\cite{Elhatisari:2022zrb} and the corresponding LECs have been fine-tuned in~\cite{Hildenbrand:2025voq}
to exactly reproduce the $^{100}$Sn binding energy. No such fine-tuning is done here.

For ground-state energy calculations, each combination of two-and three-nucleon forces has an energy minimum within the
family of nonperturbative Hamiltonians $H_s$, the so-called simple Hamiltonian, obtained by varying the two-body contact strength $C_2$. In the WFM framework, observables for the full Hamiltonian $H$ are calculated by treating the difference $H-H_s$ perturbatively, substantially suppressing sign oscillations.
We perform all production calculations with the simple
Hamiltonian used for the original WFM interaction. To assess the suitability of this choice as we vary the three-nucleon interaction, we scan a wide range of $C_2$ values for
$^{100}$Sn at $L_t=1200$, where the energy is nearly converged and the phase remains favorable. We include an interaction in the analysis if its energy at the production value of $C_2$ lies within $1\%$ of the minimum found in this scan. This relative tolerance scales with the binding energy. Approximately $41\%$ of the interactions satisfy
this simple Hamiltonian selection criterion.
This allows us to investigate correlations between binding
energies and radii in the heavy-mass region. The stochastic search combined with optimization in
Ref.~\cite{Elhatisari:2026dzq} produces an extensive list of interactions ranked by their root-mean-square deviation (RMSD) from nuclear binding-energy and neutron--alpha scattering data. Here, we keep the operator combinations and fitted low-energy constants unchanged and extend the ranking score to include the tin binding energies. For each interaction satisfying the simple Hamiltonian selection criterion, we compute a modified RMSD$_{\text{mod}}$ from the original RMSD$_{\text{fit}}$,
\begin{equation}\label{eq: fitfolding}
    \text{RMSD}_{\text{mod}}=\sqrt{\frac{N_F \text{RMSD}_{\text{fit}}^2+\sum_i d_i^2}{N_F+i_{\text{max}}}}\,,
\end{equation}
in which $N_F$ is the number of nuclei and scattering data included in the original fit, $i_{\text{max}}$ is the number of included tin isotopes, and $d_i=(B_i^{\text{th}}-B_i^{\text{ref}})/A_i$ is the binding-energy residual per nucleon, expressed in MeV. The reference binding energies are taken from experimental data and the predictions of Ref.~\cite{Ireland:2025tui}. We then reorder the interactions within each category
according to RMSD$_{\text{mod}}$, without readjusting any
low-energy constants.

Since we also collect new data for the energy we can also provide two neutron separation energies $S_{2n}$. These energies are obtained from two different procedures. The most naive one is to  take $S_{2n}=B^{\tau=\infty}(Z,N)-B^{\tau=\infty}(Z,N-2)$ of previously extrapolated values.
More effectively in terms of uncertainties, is to extrapolate $S_{2n}^{_{\tau_i}}=B^{\tau_i}(Z,N)-B^{\tau_i}(Z,N-2)$ directly. Both methods agree within uncertainties, however, the difference converges much faster than the total energy and therefore allows a more reliable estimate. Indeed it is almost $L_t$ independent. However, this method  fails for \nuc{104}{Sn}.

\section{Results and discussion}
We present energy and radius calculations at $La=18.48$~fm for the tin isotopes $^{96,98,100,102,104}$Sn near the proton dripline, using the N$^3$LO two-nucleon interaction and the three-nucleon forces described above. The study includes isotopes beyond the experimentally established end of the chain, for which binding-energy estimates rely on extrapolations

\begin{figure}[htb]
    \centering
    \includegraphics[width=1.0\linewidth]{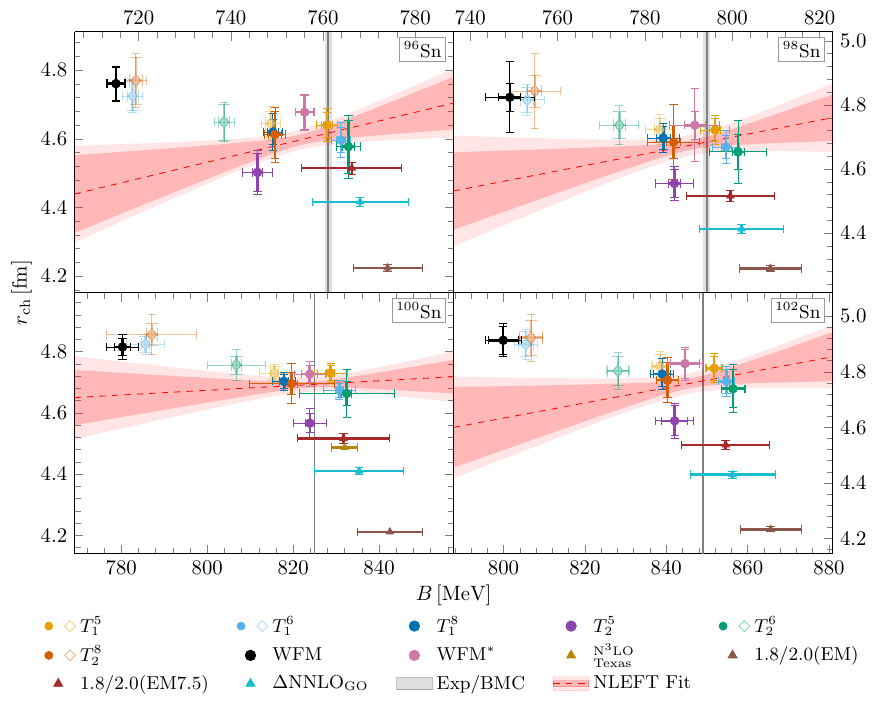}
\caption{Charge radii of $^{96,98,100,102}$Sn for different PWD three-body interactions. Open, faded symbols denote the best fits selected without tin data, and filled symbols denote the best fit in each category including tin energies. Triangles are other \emph{ab initio} calculations with different interactions. Vertical gray bands indicate experimental or BMC binding-energy references~\cite{Ireland:2025tui}. Red dashed lines show weighted fits through the selected NLEFT results, and the bands reflect the two uncertainty evaluations described in the text.}
    \label{fig:charge_radius}
\end{figure}

Figure~\ref{fig:charge_radius} shows the charge radii of $^{96,98,100,102}$Sn as functions of the binding energy for different combinations of the three-body interactions. The original WFM interaction~\cite{Elhatisari:2022zrb} is displayed by a black filled-circle, and its modified version (WFM$^{*}$) from Ref.~\cite{Hildenbrand:2025voq} provides an additional reference. The PWD three-body interactions are shown by circles.
Within each category, open, faded diamonds denote the
highest-ranked interaction according to the original
RMSD$_{\text{fit}}$, whereas filled circles denote the
highest-ranked interaction according to the modified
RMSD$_{\text{mod}}$ of Eq.~\eqref{eq: fitfolding}, which
includes the tin binding energies. The red dashed lines are weighted fits through those combinations, with uncertainty bands based on the uncertainties of the individual points. The triangles are the results of other \emph{ab initio} calculations \cite{Demol:2026glc,Hu:2025cjl}. The BCC calculations~\cite{Demol:2026glc} are performed at fixed $E_{\text{max}}$, and we estimate their energy uncertainties from the reported variation with $\hbar\omega$. For the IMSRG results~\cite{Hu:2025cjl} we use the provided $E_{\text{max}}$ dependence and extrapolate with a single exponential to the best of our knowledge. The $^{100}$Sn result of Arthuis et al.~\cite{Arthuis:2020toz} $r_{\rm ch}=4.525~\text{fm}-4.707$~fm is not shown since no corresponding binding energy is provided. The experimental binding energy values and Bayesian model combinations (BMC) based on density functional theory predictions for the pertinent binding energies~\cite{Ireland:2025tui} are shown by the vertical gray bands.

In addition we show two error bars. The solid thick bars include Monte Carlo statistical uncertainties and uncertainties from the main Euclidean-time extrapolation, which uses the double exponential ansatz,
\begin{equation}
    \langle \mathcal{O}(\tau)\rangle=\langle\mathcal{O}(\infty)\rangle+a \exp\left(-\Delta E\tau \right)
    +b \exp\left(-{\Delta E\tau}/{2}\right)\,,
\end{equation}
where $a, b$ and $\Delta E$ are fit parameters.
These uncertainties also account for changing the order in which contributions are summed and extrapolated. The thinner error bars correspond to a fit restricted to larger Euclidean times, where excited-state contributions are suppressed, using a single-exponential ansatz. The particularly noisy $Q^4$ contributions to the radii are extrapolated separately.

\begin{figure}[htb]
    \centering
    \includegraphics[width=1.0\linewidth]{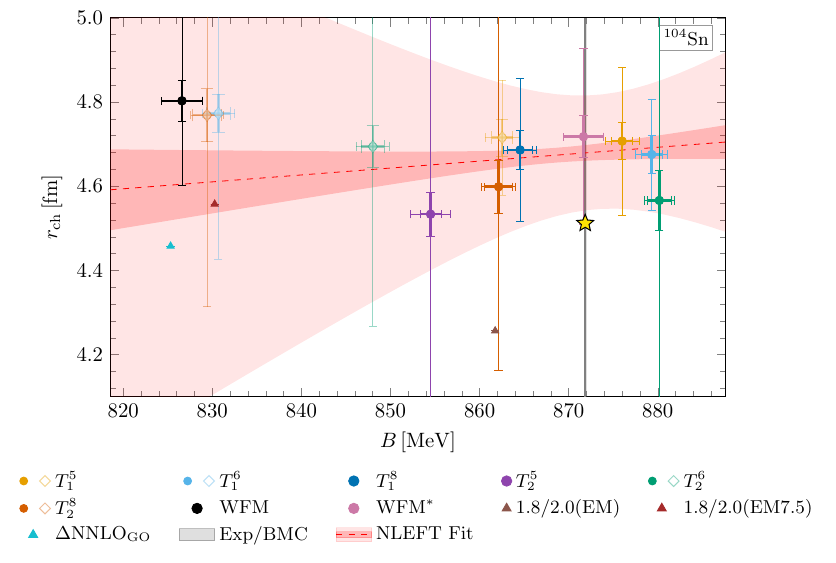}
    \caption{Charge radii of \nuc{104}{Sn} for different PWD three-body interactions, with the same interaction labels as in Fig.~\ref{fig:charge_radius}. The star denotes the experimental result~\cite{Gustafsson:2025xpd}. Unlike the comparison results for the lighter isotopes, the values from Demol et al.~\cite{Demol:2026glc} shown here do not include triples corrections or an energy uncertainty from variation of $\hbar\omega$.}
    \label{fig:Sn104}
\end{figure}

In Fig.~\ref{fig:Sn104} we show the charge-radius results for \nuc{104}{Sn}, allowing a direct comparison with experiment~\cite{Gustafsson:2025xpd}. Unlike \nuc{100}{Sn}, for which sign oscillations do not restrict the Euclidean-time range needed here, \nuc{104}{Sn} has a usable window of roughly $\tau\leq0.8$~MeV$^{-1}$, leading to substantially larger extrapolation uncertainties. The contrast with \nuc{96}{Sn}, which has the same $|N-Z|$, suggests sensitivity to the different valence-shell structures across $N=50$. In particular, mixing between neutron configurations involving the closely spaced $g_{7/2}$ and
$d_{5/2}$ orbitals above the shell closure~\cite{Darby:2010xp}
may increase the projection time required for convergence. The central values of the radius estimates in Tables~\ref{tab:res} and~\ref{tab:bay} lie above the experimental value~\cite{Gustafsson:2025xpd}, but their uncertainties limit the precision of this comparison. 

Table~\ref{tab:res} gives the radii obtained by evaluating the weighted radius--energy fits at the experimental binding energies or adopted BMC estimates. 

\renewcommand{\arraystretch}{1.1}
\begin{table}[htb]
    \sisetup{table-alignment=center, separate-uncertainty, table-align-uncertainty}
    \begin{center}
    \caption{charge  and matter radii of selected Sn isotopes at N$^3$LO for a fit trough different sets of PWD three-body forces evaluated at the experimental binding energy or BMC estimate. The given errors include statistical as well as extrapolation model uncertainties.}
 \begin{tabular}{l @{\quad} S[table-format=3.1+-2.2] @{\quad} S[table-format=1.2+-1.2] @{\quad} S[table-format=1.2+-1.2]}
\toprule
{Nucleus}& {$B$~[MeV]} & {$r_{\rm ch}$~[fm]} & {$r_m$~[fm]} \\
        \midrule
        \nuc{\phantom{1}96}{Sn} & {761.11} & 4.62 +- 0.04 & 4.51 +- 0.04 \\
        \nuc{\phantom{1}98}{Sn}  & {794.16} & 4.68 +- 0.04 & 4.54 +- 0.04 \\
        \nuc{100}{Sn} & {825.30} & 4.69 +- 0.03 & 4.58 +- 0.03 \\
        \nuc{102}{Sn} & {849.09} & 4.77 +- 0.04 & 4.60 +- 0.03 \\
        \nuc{104}{Sn}  & {871.85} & 4.68 +- 0.14 & 4.58 +- 0.05 \\
        \bottomrule
    \end{tabular}
    \label{tab:res}
    \end{center}
\end{table}

We note that the NLEFT charge radii are generally larger than the other {\em ab initio} predictions, although the calculations use different interactions, particularly in the three-nucleon sector. This comparison therefore probes both interaction and many-body-method dependence. Most of the considered interactions yield binding energies below the best estimate from experiments. This is, however, not true of all interactions, but rather an effect of the simple Hamiltonian selection criterion.

We also analyze all combinations of three-body forces, assigning weights,
\begin{equation}
    w_i=\frac{1}{n_{\rm cat}(i)}\int\int\dd{\ln (s)}\dd{\ln (\xi)}\mathcal{L}_i(s,\xi)\,
\end{equation}
where $s$ is the independent scatter of deviations in percentage of the binding energy, $\xi$ is a common shift in percent, while $n_{\rm cat}$ is the number of combinations in each category $T_i^p$. We integrate $s$ as well as $\xi$ out with log-uniform priors between $0.001$ and $0.5$. The likelihood $\mathcal{L}_i$ includes residuals for the original nuclear and scattering data and for the tin isotopes. Table~\ref{tab:bay} gives the resulting ensemble-weighted binding energies and radii.

\renewcommand{\arraystretch}{1.1}
    \begin{table}[htb]
        \sisetup{table-alignment=center, separate-uncertainty, table-align-uncertainty}
        \begin{center}
        \caption{Ground state binding energy, charge and matter radii of selected Sn isotopes at N$^3$LO for a Bayesian style analysis including correlated model uncertainties of the pertinent three-body forces.}
     \begin{tabular}{l  @{\quad} S[table-format=3.1+-2.1] @{\quad} S[table-format=1.2+-1.2] @{\quad} S[table-format=1.2+-1.2]}
            \toprule
{Nucleus} & {$B$~[MeV]} & {$r_{ch}$~[fm]} & {$r_m$~[fm]} \\
            \midrule
    \nuc{\phantom{1}96}{Sn}  & 752.4 +- 7.5  & 4.65 +- 0.07 & 4.53 +- 0.06 \\
    \nuc{\phantom{1}98}{Sn}  & 787.5 +- 7.9  & 4.72 +- 0.07 & 4.59 +- 0.06 \\
    \nuc{100}{Sn}            & 820.2 +- 8.1  & 4.73 +- 0.05 & 4.62 +- 0.06 \\
    \nuc{102}{Sn}            & 842.8 +- 8.0  & 4.82 +- 0.07 & 4.66 +- 0.06 \\
    \nuc{104}{Sn}            & 867.0 +- 8.2  & 4.72 +- 0.16 & 4.62 +- 0.07 \\
            \bottomrule
        \end{tabular}
        \label{tab:bay}
        \end{center}
    \end{table}

Contrary to previous radius calculations using the partial-pinhole method, in the work presented here Galilean invariance restoration (GIR) corrections of the WFM terms are fully included. These decrease the radii by approximately $\sim 0.08$~fm in the region of $A\sim100$. Together with calculations for \nuc{16}{O} and \nuc{40}{Ca}, the results suggest an approximately linear increase with  $A$ over the nuclei examined.

\begin{figure}[htb]
    \centering
    \includegraphics[width=1.0\linewidth]{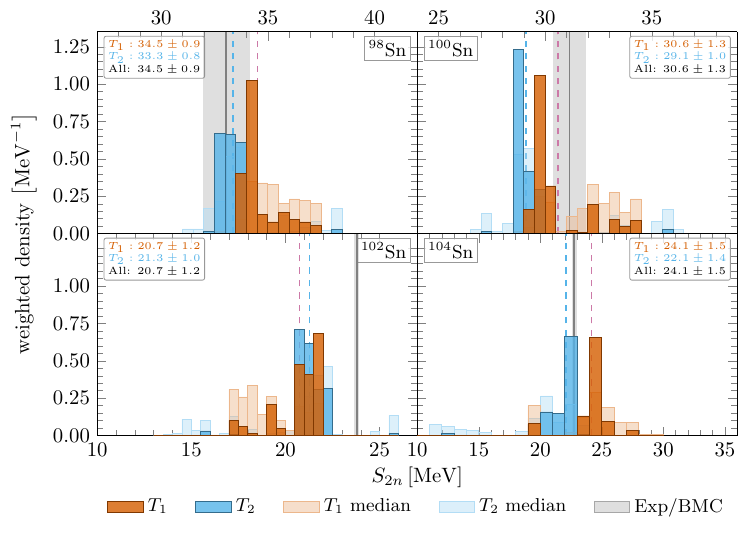}
        \caption{Separation energy for different Sn isotopes, separately shown for the two sets of three-body forces. The gray band in the experimental reference, the dashed lines corresponds to the central for $T_1$ and $T_2$. The light bars correspond to a less restrictive model based on the median of the RMSD.}
    \label{fig:sep_energies}
\end{figure}

Finally, in Fig.~\ref{fig:sep_energies} we show  the two-neutron separation energies, $S_{2n}$, calculated alongside the radii. For $^{98,100,102}$Sn, we extrapolate the finite-time energy differences, as described above, to obtain a reliable result. The separation-energy analysis uses weights based on RMSD$_{\rm mod}$. We show both sets weighted separately, for the combined sets $T_1$ carries all the weight. The shaded bars show a less restrictive version, in which we instead of integrating out $s$ and $\xi$ use a fixed error scale based on the median of RMSD$_{\text{mod}}$ over all pertinent data. Even for such a model $T_2$ would only attribute for ~2.5\% of the weight. The separation energies would be stable within $1.5$~MeV. The average separation energies and the corresponding spread are given in the panels in Fig.~\ref{fig:sep_energies}. The decrease from approximately $31$ to $20$~MeV across $N=50$ is consistent with a shell closure.

To conclude the discussion of the results, we want to make some comments of the robustness of the results. First, tightening the $C_2$ check from $1\%$ to $0.5\%$ changes the reported energies by at most $0.3$~MeV and radii by at most $0.02$~fm.  This also gives a good indication of the residual $C_2$ dependence of the results. Second, we estimate finite volume corrections to be small. We find that at large Euclidean times, the population of the most distant sampled lattice sites is nearly constant in $\tau$ and does not exceed $0.25\%$. In addition, a test calculation at $La=21.12$~fm reveals that less than $0.05\%$ would propagate to a site further away from the centre-of-mass.

\section{Summary and outlook}
We calculated charge and matter radii near the proton dripline for tin isotopes \nuc{96}{Sn}, \nuc{98}{Sn}, \nuc{100}{Sn}, \nuc{102}{Sn}, \nuc{104}{Sn} in the framework of NLEFT with high-fidelity interactions at N$^3$LO. We performed an extensive analysis of the impact of three-body forces on the radii. In addition we include GIR corrections for the WFM contributions. While their impact on light nuclei is small, the magnitude of
these contributions increases approximately linearly with $A$ over the nuclei examined, faster than the phenomenological scaling of the radius itself, $\sim A^{1/3}$. While our radii are compatible with the results from Ref.~\cite{Arthuis:2020toz}
based on self-consistent Greens function theory combined with chiral NNLO$_{\rm sat}$ forces, they are larger than the values obtained using the widely applied EM type interactions combined with BCC techniques~\cite{Demol:2026glc}. The central value of the charge radius for \nuc{104}{Sn} lies above the experimental value, but its uncertainty is large.

This work highlights the sensitivity to three-body forces in the heavy-mass region. It
underlines the observation that constraining the LECs with light- and medium-mass nuclei and scattering data does not ensure high precision for heavy nuclei. The framework
of Ref.~\cite{Elhatisari:2026dzq} allows heavy-nucleus information to be incorporated alongside scattering data. Here, we use the
available tin binding-energy information to rank existing interactions without readjusting their LECs.

\vfill
\FloatBarrier
\section{Acknowledgements}
We thank Shuang Zhang for useful discussions.
This work was supported in part by the European
Research Council (ERC) under the European Union's Horizon 2020 research
and innovation programme (grant agreement No. 101018170),
and by the CAS President's International Fellowship Initiative (PIFI) (Grant No.~2025PD0022). The work of SE is supported
in part by the Scientific and Technological Research Council of Turkey (TUBITAK project no. 123F464)
The authors thank the Gauss Centre for Supercomputing e.V. (GCS) for providing computing time on the Supercomputer JUPITER and JUWELS at the Jülich Supercomputing Centre via the John von Neumann Institute for Computing. JUPITER is supported by the EuroHPC JU and GCS through funding by the European Commission, the German Federal Ministry of Research, Technology and Space, and the Ministry of Culture and Science of the State of North Rhine-Westphalia. JUWELS is supported by GCS through funding by the German Federal Ministry of Research, Technology and Space and the Ministry of Culture and Science of the State of North Rhine-Westphalia. 
We also thank the Jülich Supercomputing Centre for providing dedicated HPC time on the \mbox{JURECA DC} GPU partition for the \mbox{EXOTIC} project.

\vfill
\pagebreak

\section{End Matter}

The two-body correlation functions on the lattice are given by
\begin{equation}\label{eq:two-body-correlation}
  G_{\alpha\beta} = \sum_{\substack{\boldsymbol{n}_1,\sigma_1,\tau_1 = \alpha \\ \boldsymbol{n}_2,\sigma_2,\tau_2 = \beta}} \langle :\rho_{\sigma_1\tau_1}(\boldsymbol{n}_1) \rho_{\sigma_2\tau_2}(\boldsymbol{n}_2): \rangle (\boldsymbol{n}_1 - \boldsymbol{n}_2)^2,
\end{equation}
where $\boldsymbol{n}$ are lattice grid coordinates and 
$\rho_{\sigma_1\tau_1}$ are the densities of particles with the corresponding spin ($\sigma$) and isospin $(\tau)$ degrees of freedom, and $::$ denotes normal ordering.

\begin{figure}[h!]
    \centering
    \includegraphics[width=1.0\linewidth]{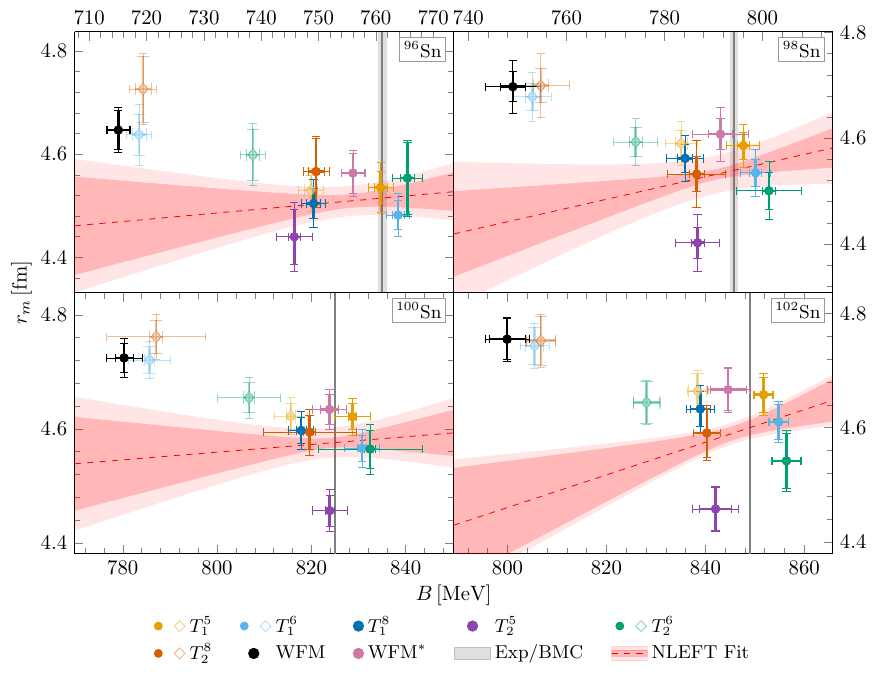}
        \caption{Matter radii calculations of tin isotopes for different PWD three-body forces, open faded symbols are the best fit without folding in the tin isotopes, full dots are the best fist per fit category with with folded in  tin isotopes. The experimental binding energy or  BMC extrapolation is given by vertical lines~\cite{Ireland:2025tui}. The red dashed line corresponds to a weighted fit of the best NLEFT interactions per category, the uncertainty bands correspond to the uncertainty methodology of the singular points.}
    \label{fig:matter_radius}
\end{figure}

In Fig.~\ref{fig:matter_radius} we show predictions for the matter radii of the same isotopes. We note that these can not be measured model-independently, cf. the corresponding discussion in~\cite{Ren:2025vpe}.

In addition to the main extrapolation results of the main text, we want to provide the results of the best fits presented in Fig.~\ref{fig:charge_radius}. The results are given in Table~\ref{tab:res_long}.

\renewcommand{\arraystretch}{1.1}
\begin{table}[htb]
    \sisetup{table-alignment=center, separate-uncertainty, table-align-uncertainty}
    \begin{center}
    \caption{Ground state binding energy, charge radius and matter radius of selected Sn isotopes at N$^3$LO for different sets of partial wave decomposed three-body forces. Errors are the total uncertainties.}
    \begin{tabular}{l c @{\quad} S[table-format=3.1+-2.1] @{\quad} S[table-format=1.2+-1.2] @{\quad} S[table-format=1.2+-1.2]}
   \toprule
        {Nucleus} & {TBF} & {$B$~[MeV]} & {$r_{ch}$~[fm]} & {$r_m$~[fm]} \\
        \midrule
        \nuc{96}{Sn} & $T_1^5$ & 760.8 +- 2.3 & 4.64 +- 0.05 & 4.54 +- 0.05 \\
         & $T_1^6$ & 763.9 +- 2.1 & 4.60 +- 0.06 & 4.48 +- 0.05 \\
         & $T_1^8$ & 749.1 +- 2.2 & 4.62 +- 0.06 & 4.50 +- 0.05 \\
         & $T_2^5$ & 745.8 +- 3.3 & 4.50 +- 0.07 & 4.44 +- 0.07 \\
         & $T_2^6$ & 765.5 +- 2.6 & 4.58 +- 0.10 & 4.55 +- 0.08 \\
         & $T_2^8$ & 749.6 +- 2.4 & 4.61 +- 0.09 & 4.57 +- 0.07 \\
        \midrule
        \nuc{98}{Sn} & $T_1^5$ & 796.1 +- 3.4 & 4.72 +- 0.05 & 4.59 +- 0.05 \\
         & $T_1^6$ & 798.6 +- 3.2 & 4.67 +- 0.06 & 4.53 +- 0.05 \\
         & $T_1^8$ & 784.2 +- 3.8 & 4.70 +- 0.05 & 4.56 +- 0.05 \\
         & $T_2^5$ & 786.8 +- 4.5 & 4.56 +- 0.06 & 4.40 +- 0.06 \\
         & $T_2^6$ & 801.3 +- 6.6 & 4.65 +- 0.10 & 4.50 +- 0.06 \\
         & $T_2^8$ & 786.6 +- 5.9 & 4.68 +- 0.12 & 4.53 +- 0.07 \\
        \midrule
        \nuc{100}{Sn} & $T_1^5$ & 828.7 +- 3.8 & 4.73 +- 0.03 & 4.62 +- 0.04 \\
         & $T_1^6$ & 830.7 +- 3.7 & 4.67 +- 0.04 & 4.57 +- 0.04 \\
         & $T_1^8$ & 817.9 +- 2.7 & 4.70 +- 0.03 & 4.60 +- 0.04 \\
         & $T_2^5$ & 823.9 +- 3.8 & 4.57 +- 0.05 & 4.46 +- 0.04 \\
         & $T_2^6$ & 832.5 +- 11.1 & 4.66 +- 0.08 & 4.56 +- 0.05 \\
         & $T_2^8$ & 819.6 +- 9.9 & 4.70 +- 0.07 & 4.59 +- 0.05 \\
        \midrule
        \nuc{102}{Sn} & $T_1^5$ & 851.8 +- 2.1 & 4.81 +- 0.06 & 4.66 +- 0.04 \\
         & $T_1^6$ & 854.8 +- 2.1 & 4.77 +- 0.06 & 4.61 +- 0.04 \\
         & $T_1^8$ & 839.0 +- 2.8 & 4.79 +- 0.06 & 4.63 +- 0.05 \\
         & $T_2^5$ & 842.1 +- 4.8 & 4.62 +- 0.07 & 4.46 +- 0.05 \\
         & $T_2^6$ & 856.4 +- 3.0 & 4.74 +- 0.09 & 4.54 +- 0.06 \\
         & $T_2^8$ & 840.3 +- 2.8 & 4.77 +- 0.09 & 4.59 +- 0.05 \\
        \midrule
        \nuc{104}{Sn} & $T_1^5$ & 876.0 +- 2.0 & 4.71 +- 0.18 & 4.61 +- 0.07 \\
         & $T_1^6$ & 879.3 +- 1.8 & 4.68 +- 0.14 & 4.57 +- 0.05 \\
         & $T_1^8$ & 864.5 +- 1.9 & 4.69 +- 0.17 & 4.58 +- 0.06 \\
         & $T_2^5$ & 854.5 +- 2.3 & 4.53 +- 0.84 & 4.45 +- 0.37 \\
         & $T_2^6$ & 880.2 +- 1.7 & 4.57 +- 0.69 & 4.47 +- 0.41 \\
         & $T_2^8$ & 862.1 +- 1.9 & 4.60 +- 0.44 & 4.52 +- 0.28 \\
        \bottomrule

    \end{tabular}
    \label{tab:res_long}
    \end{center}
\end{table}

\clearpage

\bibliographystyle{unsrturl}

\bibliography{reference}

\end{document}